\documentclass[sigconf]{acmart}

\AtBeginDocument{%
  \providecommand\BibTeX{{%
    \normalfont B\kern-0.5em{\scshape i\kern-0.25em b}\kern-0.8em\TeX}}}

\setcopyright{acmcopyright}
\copyrightyear{2021}
\acmYear{2021}
\acmDOI{10.1145/1122445.1122456}

\copyrightyear{2021} 
\acmYear{2021} 
\setcopyright{rightsretained} 
\acmConference[CHI PLAY '21]{Extended Abstracts of the 2021 Annual Symposium on Computer-Human Interaction in Play}{October 18--21, 2021}{Virtual Event, Austria}
\acmBooktitle{Extended Abstracts of the 2021 Annual Symposium on Computer-Human Interaction in Play (CHI PLAY '21), October 18--21, 2021, Virtual Event, Austria}\acmDOI{10.1145/3450337.3483505}
\acmISBN{978-1-4503-8356-1/21/10}

\begin{document}

\title{Myopic Bike and Say Hi: Games for Empathizing with The Myopic}


\author{Xiang Li}
\affiliation{%
  \institution{Xi'an Jiaotong-Liverpool University}
  \city{Suzhou}
  \country{China}
}
\affiliation{%
  \institution{Exertion Games Lab\\ Monash University}
  \city{Melbourne}
  \country{Australia}
}
\email{xiang@exertiongameslab.org}

\author{Xiaohang Tang}
\affiliation{%
  \institution{Xi'an Jiaotong-Liverpool University}
  \city{Suzhou}
  \country{China}
}
\email{xiaohang.tang19@student.xjtlu.edu.cn}

\author{Xin Tong}
\affiliation{%
  \institution{Duke Kunshan University}
  \city{Suzhou}
  \country{China}
}
\affiliation{%
  \institution{Stanford University}
  \city{Stanford}
  \country{USA}
}
\email{xt43@duke.edu}

\author{Rakesh Patibanda}
\affiliation{%
  \institution{Exertion Games Lab\\ Monash University}
  \city{Melbourne}
  \country{Australia}
}
\email{rakesh@exertiongameslab.org}

\author{Florian 'Floyd' Mueller}
\affiliation{%
  \institution{Exertion Games Lab\\ Monash University}
  \city{Melbourne}
  \country{Australia}
}
\email{floyd@exertiongameslab.org}

\author{Hai-Ning Liang}
\authornote{Corresponding author: haining.liang@xjtlu.edu.cn}
\affiliation{%
  \institution{Xi'an Jiaotong-Liverpool University}
  \city{Suzhou}
  \country{China}
}
\email{haining.liang@xjtlu.edu.cn}

\renewcommand{\shortauthors}{Xiang Li et al.}

\begin{abstract}
Myopia is an eye condition that makes it difficult for people to focus on faraway objects. It has become one of the most serious eye conditions worldwide and negatively impacts the quality of life of those who suffer from it. Although myopia is prevalent, many non-myopic people have misconceptions about it and encounter challenges empathizing those who suffer from it. In this research, we developed two virtual reality (VR) games, (1) "Myopic Bike" and (2) "Say Hi", to provide a means for the non-myopic population to experience the difficulties of myopic people. Our two games simulate two inconvenient daily life scenarios (riding a bicycle and greeting friends on the street) that myopic people encounter when not wearing glasses. The goal is to facilitate empathy in people with non-myopia for those who suffer from myopia. We evaluated four participants' game experiences through questionnaires and semi-structured interviews. Overall, our two VR games can create an engaging and non-judgmental experience for the non-myopic people that has potential to facilitate empathizing with those who suffer from myopia.
\end{abstract}




\begin{teaserfigure}
    \includegraphics[width=\textwidth]{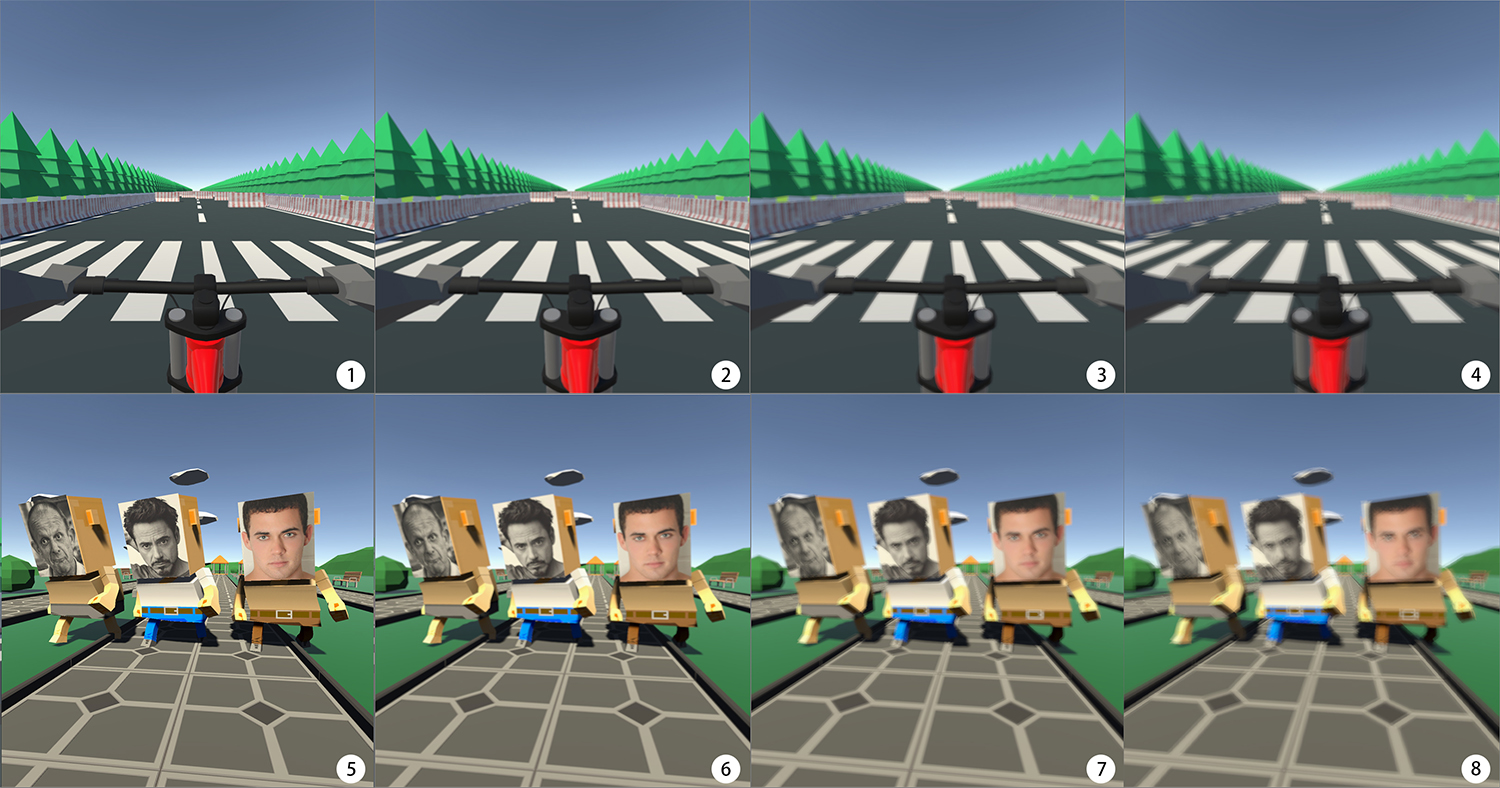}
    \centering
    \caption{Two VR games facilitate empathising people with myopia: \textit{Myopic Bike} and \textit{Say Hi}. 1-4 and 5-8 simulate four myopic conditions: no myopia, low, moderate and high myopia (from left to right).}
    \Description{Screenshots of the two games developed to help people empathise with myopic users. (Top row: 1.1 to 1.4): \textit{Myopic Bike}; (Bottom row:1.5 to 1.8): \textit{Say Hi}, in four myopic conditions: No Myopia, Low Myopia, Moderate Myopia, and High Myopia (from left to right).}
\end{teaserfigure}

\maketitle

\section{Introduction}
Myopia (or nearsightedness) is a common vision condition with which people can see nearby objects clearly but the further-away ones are blurry \cite{morgan2012myopia}. Myopia happens when the shape of an eye causes light rays to bend (refract) incorrectly, focusing images in front of the retina instead of on the retina \cite{morgan2012myopia}. In general, low myopia is less than 3.0 diopters (\textless -3.0 D), moderate myopia is less than 6.0 diopters (-3.0 D to -6.0 D), and high myopia is usually greater than 6.0 diopters (\textgreater -6.0 D). 

Myopia is one of the most common eye problems worldwide. For example, approximately 600 million residents in China are suffering from myopia \cite{zhao2018virtual} and 4\% of the population in the United States are suffering from high myopia \cite{sia2014us}. For non-myopic people, it is often difficult to empathize with different levels of myopia, and as such, it can be challenging to feel the distress and discomfort experienced by those who have it \cite{benford2013uncomfortable}. We believe that the lack of empathy toward the population with myopia could result in inadequate considerations when designing products for myopic people and inadvertently cause accessibility problems.

Recently, virtual reality (VR) games has shown great potential as a medium to foster empathy in a non-judgemental but engaging and fun way \cite{shin2018empathy}. VR can offer players an embodied first-person perspective to experience different severities of myopia throguh the use of a virtual environment \cite{arnold2018you}. Unlike traditional displays on mobile phones or monitors, VR gives the user a depth experience, bringing the most visual-based and non-tactile feedback in such non-existent scenarios \cite{li2021vrcaptcha}, which iempathy with patients experiencings beneficial for scenario simulation and immersive experience.

The use of extended reality to study health issues is not new. For example, Xu et al. \cite{xu2020results} designed a variety of VR exergames in seated or standing postures using head-mounted displays (HMDs). Likewise, Xu and colleagues \cite{xu2020virusboxing} designed a high-intensity interval training (HIIT) boxing VR game that allowed players to achieve high heart rates and exertion, rapid fat burning, and reduced exercise time with high intensity. 

In this research, we explore the design of VR games that are aimed to foster in non-myopic people empathy toward those around us who suffer from myopia. To do this, we first collected opinions from eight myopic and two non-myopic people through a survey. We then investigated and studied the potential risks encountered by myopic people when not wearing glasses. After, we developed two VR myopic games, \textit{Myopic Bike} and \textit{Say Hi} (Figure 1). We then recruited four participants with three levels of myopia to evaluate the two games. We collected their feedback through semi-structured interviews and questionnaires. These findings suggested that these two VR games represent a good balance between having an engaging game experience and developing empathetic feelings toward people with myopia. Overall, the results of our research suggest that VR games are a promising approach to foster empathy toward myopic people.

\section{Related Work}

\subsection{Facilitating Empathy}
One common definition of empathy is "the ability to perceive, understand, and respond to the experiences and behaviors of others" \cite{szanto2019introduction}. According to Davis \cite{davis2018empathy}, empathy is manifested in three aspects: (1) physical sensations, (2) emotional sensations, and (3) cognitive awareness. As VR can provide users with an embodied and immersive experience \cite{bowman2007virtual}, it is a suitable and powerful platform to foster empathy toward those with physical and visual disorders or challenges \cite{freina2015immersive}.

Researchers have been using videogames and VR environments to foster empathy toward specific populations \cite{konrath2011changes, kral2018neural, greitemeyer2010playing, sterkenburg2019effect}. Their findings suggest that VR games can elicit a strong sense of embodied empathy because of VR's immersive nature \cite{shin2018empathy, patibanda2017life, floyd2021limited, perez2018virtual}. Furthermore, the immersive features of VR games can also provide benefits to individuals' perceptions and behaviors \cite{herrera2018building,slater2016enhancing, lim2016just, boltz2015rethinking}. Bailenson et al. \cite{herrera2018building} found that two experiential aspects of VR, immersion and embodiment, play an essential role in fostering empathy. Similarly, Tong and colleagues \cite{jin2016if, tong2020designing} designed a VR game named AS IF, and found that it promoted empathy toward chronic pain patients through an embodied avatar and game tasks. They identified the critical role of embodiment in eliciting people's empathetic attitudes. Although previous studies have explored empathy toward vulnerable populations, our study targets a unique group, people with different levels of myopia.

\subsection{Accessibility Designs for Visual Impairments}
\begin{figure*}[h]
  \centering
  \includegraphics[width=0.8\linewidth]{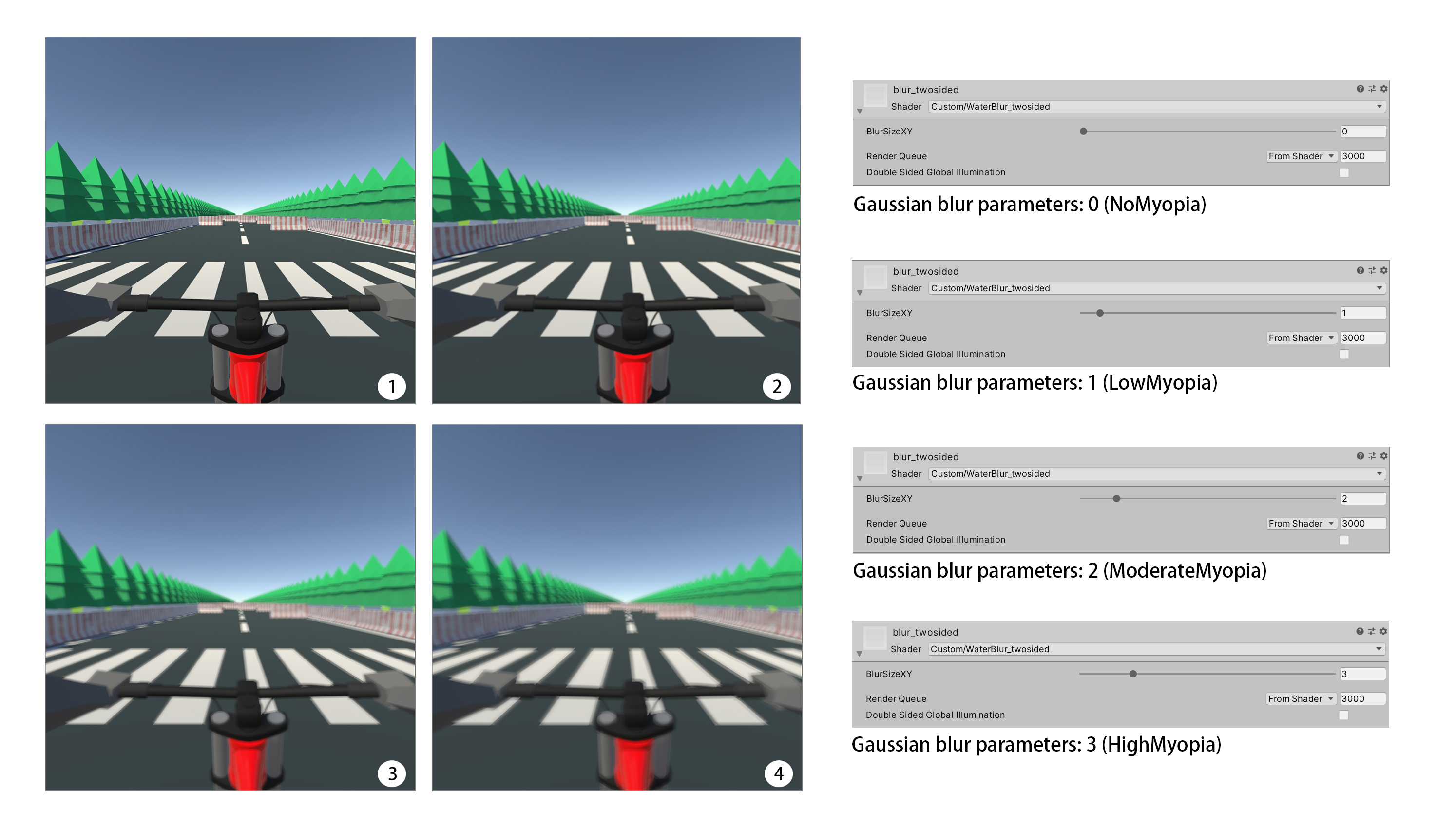}
  \caption{Four conditions in Myopic Bike. We engaged three groups of users with different degrees of myopia (low, moderate and high myopia), two from each group, in a pre-experiment data collection survey to help frame the design of the games and, based on their feedback, we adjusted the Gaussian blur parameters to match the real-world myopic situations as much as possible, which are 0, 1, 2 and 3 (Figures 2.1 - 2.4, respectively).}
  \Description{Four conditions in Myopic Bike. We engaged three groups of users with different degrees of myopia (low, moderate and high myopia), two from each group, in a pre-experiment data collection survey to help frame the design of the games and, based on their feedback, we adjusted the Gaussian blur parameters to match the real-world myopic situations as much as possible, which are 0, 1, 2 and 3 (Figures 2.1 - 2.4, respectively).}
\end{figure*}

Prior research on the accessibility design for those with visual impairments in VR has focused on blindness \cite{gonzalez2006seeing,maidenbaum2013increasing,picinali2014exploration,sanchez1999virtual}, guidance systems \cite{seki2010training,torres2010applications,zhao2019seeingvr,zhao2019designing,lecuyer2003homere,schloerb2010blindaid,tzovaras2009interactive,tzovaras2002design}, and non-visual interfaces \cite{an2020decoding,shinohara2011shadow}. Researchers also explored haptic interaction techniques for people with visual impairments \cite{siu2020virtual,zhao2018enabling,jansson1999haptic}. Krosl et al. \cite{krosl2020cataract} simulated the perspective of a person suffering from cataracts in AR to foster empathy with patients experiencing. Prior work contributed techniques that are useful for blind people or low vision populations. Our two games focus on fostering empathy toward people with severe myopia from those without myopia or with low or moderate myopic conditions. We believe that it is only by truly experiencing simulated difficulties of varying degrees of myopia that we can truly achieve empathy and thus have a deeper awareness of their challenges and ultimately provide improved accessibility designs for myopic users.

\section{Myopic Games}
\subsection{Background}
Benford et al. introduced the concept of "uncomfortable interaction" \cite{benford2012uncomfortable, benford2013uncomfortable}, which explores the benefits and unexplored potential interests of uncomfortable user experience in games. Our work falls under the category of uncomfortable interaction from the perspective of the players' experience to deliver an entertaining and fun but also enlightening and socially bonding experience \cite{benford2013uncomfortable}, specifically in the context of experiencing how myopic people feel about doing common daily activities.

We first collected myopic people's challenges in their daily lives and non-myopic people's perceptions of these challenges through an online survey. We received a total of 10 responses (6 males and 4 females with an average age of 27 years), 8 from myopic people (7 with moderate myopia in both eyes and 1 with low myopia in both eyes) and 2 responses from non-myopic people. 

Nine participants had prior some experience with VR devices. Eight myopic people reported that if they forget to wear glasses in their daily lives, they would be afraid of riding a bicycle because they could not see any obstacles on their way. Six thought they would be afraid of greeting people on the street because they could not recognize them. The other two non-myopic respondents also mentioned that they thought myopic people would be afraid to use transportation and greet people because of any embarrassing situation if they could not recognize them. Based on these initial findings, we decided to design the two myopia games, one for each scenario, i.e., \textit{Myopic Bike} and \textit{Say Hi}.

\subsection{Innovations}
Our two games have three innovative features. First, our targeted audience are people and people with low to medium myopia. Second, we provided four game levels, each mapped to a level of myopia, for players to choose from to experience different degrees of myopic situations. Third, our two VR games simulated challenges people with myopia encounter in their daily life.

\subsection{Myopic Bike}
\begin{table*}[ht]
\resizebox{\linewidth}{16mm}{
\begin{tabular}{|l|c|c|c|c|c|c|c|c|}
\hline
\multicolumn{1}{|c|}{}                     & \multicolumn{4}{c|}{\textbf{Myopic Bike}}                              & \multicolumn{4}{c|}{\textbf{Say Hi}}                                    \\ \hline
\multicolumn{1}{|c|}{\textit{}}            & NoMyopia        & LowMyopia       & ModerateMyopia  & HighMyopia       & NoMyopia         & LowMyopia       & ModerateMyopia  & HighMyopia       \\ \hline
Competence                                 & 2.65 (.44) & 2.65 (.79) & 2.50 (.48) & 2.70 (.42)  & 2.25 (.87)  & 1.95 (.10) & 1.85 (.77) & 1.60 (.40)  \\ \hline
\textit{Sensory and Imaginative Immersion} & 2.58 (.22)  & 2.83 (.19) & 2.67 (.30) & 2.75 (.22)  & 3.04 (.55)  & 2.71 (.25) & 2.28 (.35) & 2.79 (.50)  \\ \hline
\textit{Flow}                              & 2.25 (.19) & 2.40 (.49) & 2.30 (.26) & 2.25 (.30)  & 2.10 (.62)  & 2.05 (.57) & 2.25 (.70) & 2.05 (.64)  \\ \hline
\textit{Tension}                           & 1.05 (.66) & .80 (.43)  & 1.00 (.52) & 1.10 (.42)  & .90 (.50)   & .85 (.55)  & 1.15 (.25) & .85 (.44)   \\ \hline
\textit{Challenge}                         & 2.05 (.79) & 1.75 (.94) & 2.75 (.44) & 2.60 (1.21) & 1.80 (1.06) & 2.45 (.44) & 2.80 (.37) & 2.65 (1.25) \\ \hline
\textit{Negative affect}                   & .90 (.74)  & .90 (.70)  & 1.20 (.82) & 1.40 (.91)  & 1.50 (.50)  & 1.10 (.35) & 1.05 (.44) & 1.55 (.98)  \\ \hline
\textit{Positive affect}                   & 2.60 (.28) & 2.95 (.72) & 2.30 (.26) & 2.20 (.59)  & 2.65 (.41)  & 2.15 (.44) & 1.95 (.19) & 2.00 (.40)  \\ \hline
\end{tabular}}
\caption{Players' component mean (SD) scores for the seven core GEQ modules. }
\Description{Players' component mean (SD) scores for the seven core GEQ modules.}
\end{table*}

\begin{figure}[ht]
  \centering
  \includegraphics[width=\linewidth]{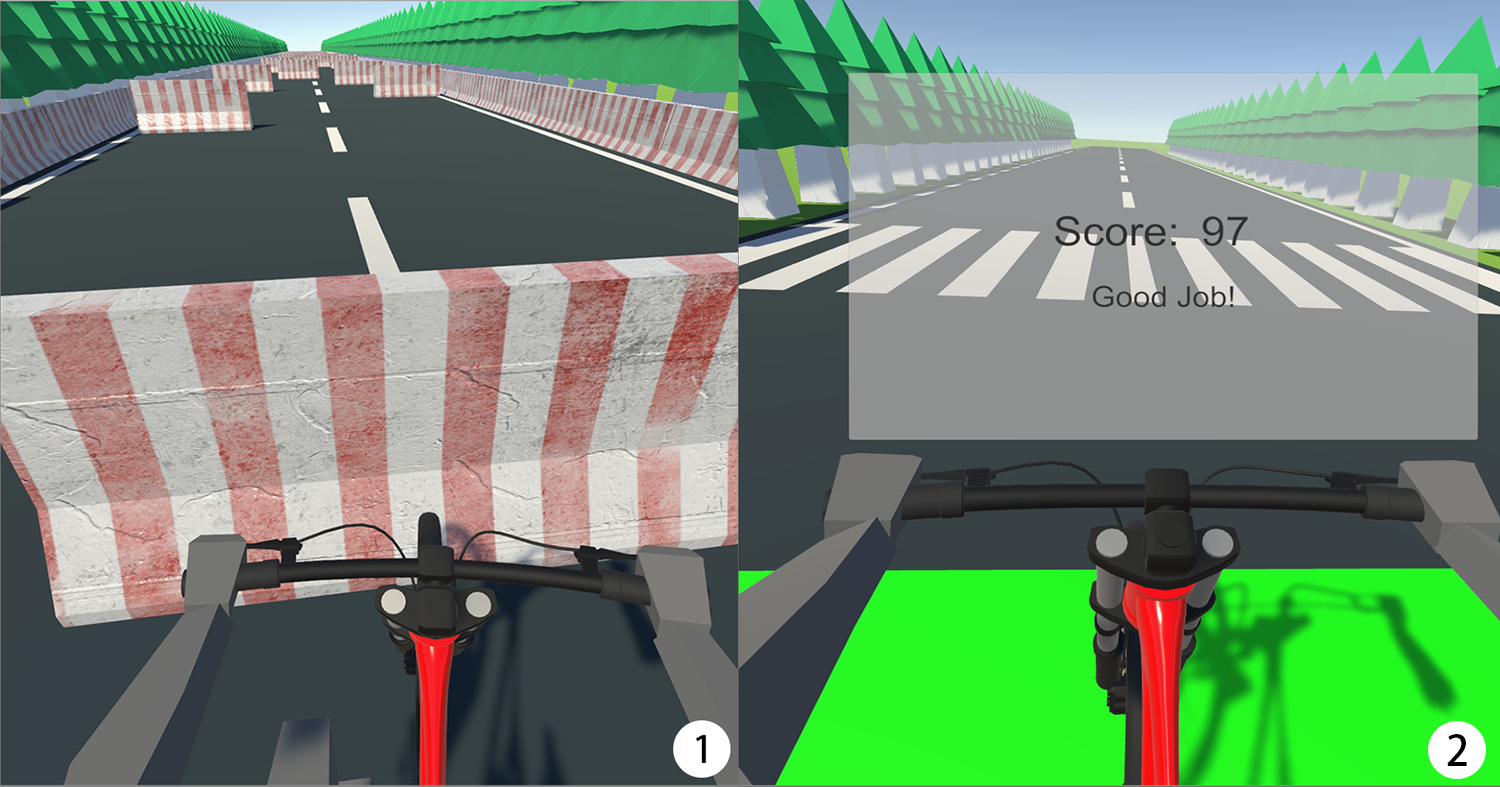}
  \caption{Myopic Bike showing (1) Roadblocks and (2) the Scoreboard (current Game Score = 97).}
  \Description{Screenshots of Myopic Bike showing (1) Roadblocks and (2) the Scoreboard.}
\end{figure}

\subsubsection{Game Concepts and Gameplay Mechanics}
\textit{Myopic Bike} offers four levels of myopia for players to choose from (NoMyopia, LowMyopia, ModerateMyopia and HighMyopia) (Fig. 2). Users play a person with myopia riding a bike. Players need to reach the end of the route with roadblocks (Fig. 3.1). There are three levels of difficulty for each of the four myopia conditions according to the number of roadblocks along the route (i.e., 40, 60, 80). Thus, players need to be skilled at handling and maneuvering the bike to avoid hitting the roadblocks and to not losing points. We also provide sound effects when a roadblock is hit and a scoreboard (Fig. 3.2) to show players their performance when they cross the finish line.

\subsection{Say Hi}

\begin{figure}[h]
  \centering
  \includegraphics[width=\linewidth]{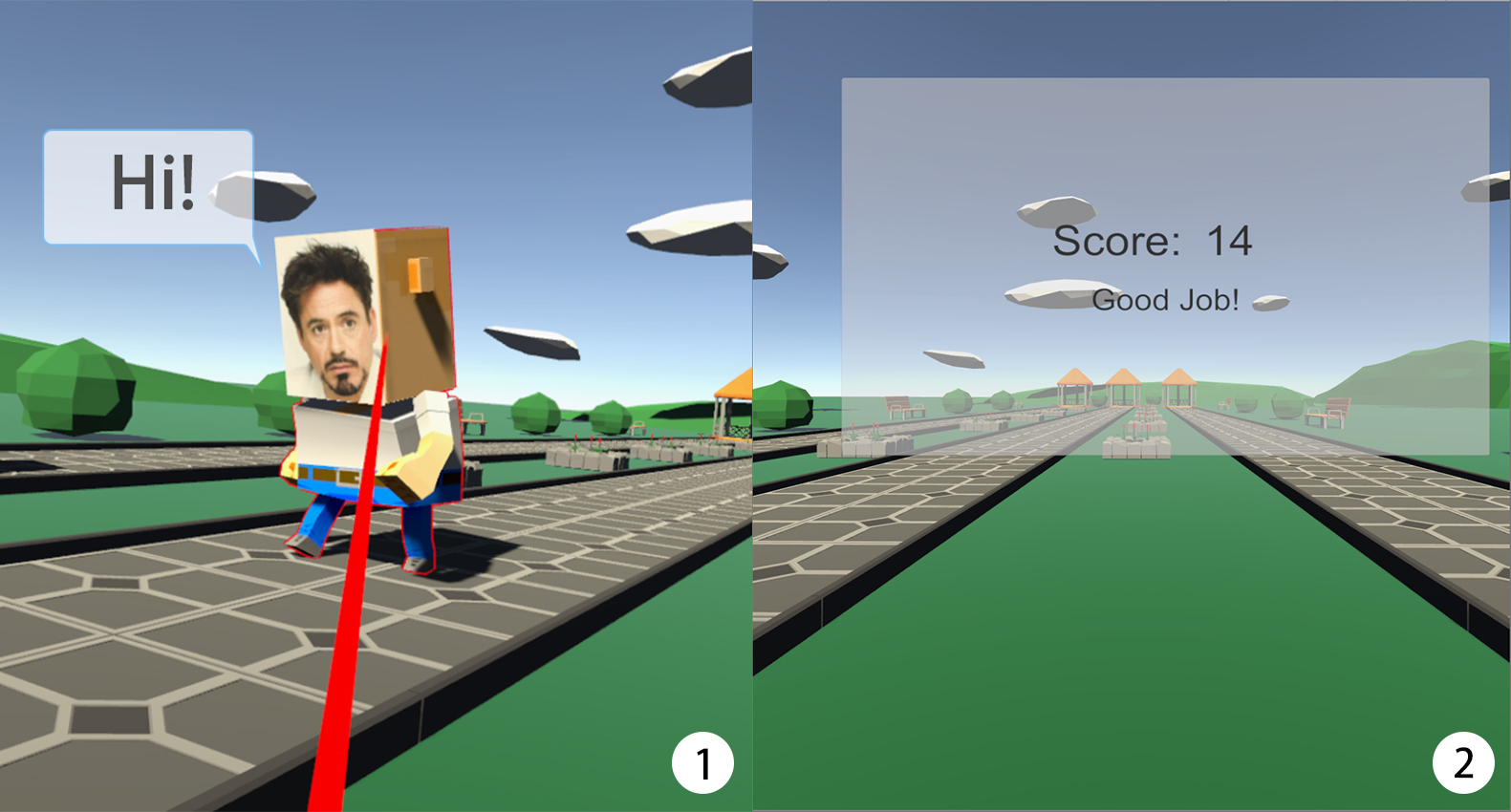}
  \caption{Screenshots of Say Hi showing (1) a virtual character to say hi to (or not), and (2) the scoreboard (current game score = 14).}
  \Description{Screenshots of Say Hi showing (1) a virtual character to say hi to (or not), and (2) the scoreboard (current game score = 14).}
\end{figure}

\subsubsection{Overall Concept and Gameplay Mechanics}
In \textit{Say Hi}, players take on the role of a pedestrian standing still and greeting "acquaintances" who are among other pedestrians. We use four international famous people as "acquaintances" for players to choose. Before starting the game, we show their pictures to the participants to ensure they know these faces. Groups of pedestrians will walk facing the player at the same speed, and disappear after 15 seconds. In each group, the game will show either one or no person from the four faces. Players need to select the one they know from all pedestrians before they disappear (Fig. 4.1). The selected pedestrian will provide positive or negative audio feedback after the player made a choice. Players' performances will be shown on a scoreboard (Fig. 4.2) when they complete the game.

We predefined the number of groups in each round and the number of groups without any acquaintance (i.e., 20 and 4 in our game). Moreover, for each group, pedestrians' clothes and faces are randomly allocated. There are three difficulty levels , the higher the level, the more pedestrians are on the road (i.e., 1 per level in our game).

\subsection{Development and Apparatus}
We used the Unity\footnote{Unity: \url{https://unity.com/}} to develop the two myopic games and Oculus Rift S\footnote{Oculus Rift S: \url{https://www.oculus.com/rift-s/}} as the VR HMD. All development and evaluation were conducted at a university's research lab.

\subsection{Playtesting}

\begin{table}[ht]
\resizebox{\linewidth}{6mm}{
\begin{tabular}{|c|c|c|c|c|}
\hline
                     & \textbf{NoMyopia} & \textbf{LowMyopia} & \textbf{ModerateMyopia} & \textbf{HighMyopia} \\ \hline
\textit{Myopic Bike} & 0.25              & 2                  & 2.75                    & 4                   \\ \hline
\textit{Say Hi}      & 0.25              & 2                  & 3.25                    & 4                   \\ \hline
\end{tabular}}
\caption{Players' empathy in the two games with different myopic conditions collected via 5-likert scale questions, from 0 to 4.}
\Description{Players' empathy in the two games with different myopic conditions collected via 5-likert scale questions, from 0 to 4.}
\end{table}

We recruited four participants from a local university, each from a unique condition of myopia (including the non-myopic condition), i.e., NoMyopia, LowMyopia, ModerateMyopia, and HighMyopia. We collected their feedback from pre-prepared questionnaires, i.e., Game Experience Questionnaire (GEQ). Although some studies have questioned the validity of the GEQ, we hope our results still depict interesting feedback on our games \cite{law2018systematic}. \cite{ijsselsteijn2013game} (Table 1), a subjective emotion questionnaire (SEQ) using 5-Likert scale questions (Table 2), and conducted semi-structured interviews to analyze the players' preferences and empathy in each case \cite{carey2017toward}. We only tested the low difficulty mode of both VR games and asked each participant to experience all four myopic conditions and collected the game scores. After each experimental condition, participants were asked to complete the above-mentioned questionnaires. Participants finished the two games in about thirty minutes.

All participants had fun playing the two VR myopia games, but all had their own preferences. For instance, P1 (who does not have myopia) talked about his VR experience of being a myopic person. P1 said, \textit{"I enjoyed playing both VR myopic games, [because] I'm not myopic, so this is the first time I've felt the inconveniences experienced by people with different levels of myopia in their lives. It's almost impossible for me to see anything else in high myopia condition..."}. P3 (who has moderate myopia) talked about his previous misconceptions about high myopia, saying that \textit{"I thought before that high myopia might be nothing [different] [because] I was moderately nearsighted. But after experiencing these two games I found I was wrong. These two games have enhanced my empathy [toward people with high myopia]."}

\section{Limitations and Future Work}
We acknowledge that our design work has limitations \cite{gaver2012annotated} and that there are several possibilities to extend it. Our two VR games represent myopic people's challenges through two real-life scenarios. Other scenarios are certainly possible. To help develop these, we think that body interaction theories in HCI research can complement our future work. For example, Merleau-Ponty's "motor intentionality" \cite{svanaes2013interaction}, Mueller et al.'s concept of limited body interaction \cite{floyd2021limited} and their theoretical basis about using the body as an interface for games \cite{mueller2018experiencing} could enrich our notion of body control for more immersive myopic simulations \cite{sauve2007distinguishing}. Future work can investigate how to better utilize the body as a more direct mode of interaction to enhance the myopic game experience and achieve better empathy in players.

Our study and game design can be further refined and the myopia rendering in VR could be enhanced. We only blurred the screen without using any other extra graphics rendering deformations or transformations for players to experience different cases of myopia. Moreover, moderate to high myopia is usually accompanied by other ocular diseases such as astigmatism \cite{gwiazda2000astigmatism}, which is difficult to simulate in VR. Given that our main study was focused on myopia experiences per se, future work could explore the possibility of simulating the concomitant or superimposed ocular diseases in VR.

Regarding measuring empathy, we used a self-reported questionnaire to collect empathy changes. In future work, we plan to add a way of capturing facial expressions \cite{ekman1971constants,ekman2013emotion} and EEG signals, such as Neo-Noumena \cite{semertzidis2020neo}, to record participants' affective and biological changes.

Further exploration is required to form a framework for developing myopic VR games. From the perspective of accessible design, only by fully experiencing different levels of myopia can designers effectively design accessible products for people with myopia \cite{shinohara2011shadow}. Aside from product design, the ability to empathize with people with different degrees of myopia is conducive to making our society more inclusive and understanding towards those with visual challenges. Therefore, in the future we plan to include a larger sample size to further evaluate our VR games, extract design inspirations and recommendations. Our long term plan is to formulate a framework that can guide the design of VR games aimed at fostering people's empathetic attitudes toward people with different degrees of myopia.

\section{Conclusion}
In this paper, we presented two VR myopic games, \textit{Myopic Bike} and \textit{Say Hi}, that allows players to experience the challenges people with myopia encounter in their everyday life. that myopic people encounter when not wearing glasses. We evaluated these two games with four participants and collected their feedback using questionnaires and semi-structured interviews. Findings suggested that the two VR myopic games could foster empathy toward people with myopia.

\begin{acks}
We thank all the participants for their time. We also thank the reviewers for their comments. Xiang Li would like to thank the internship program provided by the Exertion Games Lab of Monash University. Rakesh Patibanda and Florian 'Floyd' Mueller thank the Australian Research Council. This research was supported in part by Xi'an Jiaotong-Liverpool University's Key Special Fund (\#KSF-A-03).

\end{acks}

\bibliographystyle{ACM-Reference-Format}
\bibliography{sample-base}


\end{document}